# Entanglement from longitudinal and scalar photons

J.D. Franson

*Physics Department, University of Maryland, Baltimore County, Baltimore, MD 21250 USA*

The covariant quantization of the electromagnetic field in the Lorentz gauge gives rise to longitudinal and scalar photons in addition to the usual transverse photons. It is shown here that the exchange of longitudinal and scalar photons can produce entanglement between two distant atoms or harmonic oscillators. The form of the entangled states produced in this way is very different from that obtained in the Coulomb gauge, where the longitudinal and scalar photons do not exist. A generalized gauge transformation is used to show that all physically observable effects are the same in the two gauges, despite the differences in the form of the entangled states. An approach of this kind may be useful for a covariant description of the dynamics of quantum information processing.

## I. INTRODUCTION

The covariant quantization of the electromagnetic field in the Lorentz gauge involves longitudinal and scalar (temporal) photons in addition to the transverse photons that are familiar from the Coulomb gauge [1-6]. That is necessary because the vector and scalar potentials form the components of a relativistic four-vector, and all four components must be quantized in order to maintain manifest covariance. Although the Lorentz and Coulomb gauges are physically equivalent, a manifestly covariant treatment of photons may be useful in order to provide a covariant description of the generation of entanglement and of quantum information processing.

It will be shown here that the exchange of longitudinal and scalar photons can produce entanglement between two atoms or harmonic oscillators as illustrated in Fig. 1. The form of the entangled state produced in this way is very different from that obtained in the Coulomb gauge. Nevertheless, it will be explicitly shown that the results in the two gauges are physically equivalent. Simple examples of this kind provide useful insight into the way in which the two gauges are equivalent despite their apparent differences.

Most experiments demonstrating entanglement, quantum teleportation, and quantum information processing have been analyzed using a theory of photons (the Coulomb gauge) that is not manifestly covariant. The role of special relativity and covariance in entanglement and quantum information has been discussed in a number of earlier papers [7-25], none of which are based on a covariant description of the photons in the Lorentz gauge. A covariant polarization for the photons has often been used in the Coulomb gauge, which provides a correct description of entangled states and quantum information under Lorentz transformations. The use of the covariant quantization of the electromagnetic field in the Lorentz gauge goes a step further and allows a manifestly covariant description of the time evolution of the system, including a covariant form for the Hamiltonian and perturbation theory [4].

In addition to being useful for a covariant description of the dynamics of entanglement and quantum information processing, these results provide additional insight into the techniques used to quantitatively measure entanglement. In particular, the question arises as to whether or not the usual measures of entanglement would give results that are the same with or without the entanglement from the longitudinal and scalar photons.

The longitudinal and scalar photons do not physically exist in a freely-propagating beam of light, and the theory is designed in such a way that the scalar photons are associated with negative probabilities (the indefinite metric) that cancel the effects of the longitudinal photons in the absence of any interaction [1-6]. That is not the case in the presence of a charge or current distribution, such as in an atom, where the longitudinal and scalar photons can produce physically observable effects such as the Coulomb force. It should be emphasized that the inclusion of longitudinal and scalar photons along with the indefinite metric forms the basis for the accepted covariant formulation of quantum electrodynamics [1-6].

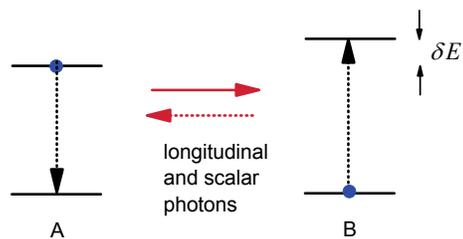

Fig. 1. Entanglement between two atoms or harmonic oscillators $A$ and $B$ produced by the exchange of virtual longitudinal and scalar photons. In the absence of any interaction, oscillator $A$ is assumed to be in its first excited state while oscillator $B$ is in its ground state. The energies of the first excited states are assumed to differ by $\delta E = \hbar \omega_B - \hbar \omega_A$. The perturbed eigenstate of the system includes a probability amplitude $\varepsilon$ for oscillator A to be in its ground state with oscillator B in its excited state, as can be calculated using perturbation theory.

Since the Lorentz gauge is not widely used in quantum optics, a brief review of the covariant quantization of the electromagnetic field in the Lorentz gauge is given in the next section. The entanglement between two harmonic oscillators produced by the exchange of longitudinal and scalar photons in the Lorentz gauge is then calculated in Section III. The corresponding situation is considered in the Coulomb gauge in Section IV, where the equivalent interaction between the harmonic oscillators is due to the classical scalar potential. The observable properties of the entangled states are then compared in Section V, where it is shown that there is no physical difference between them. A summary and conclusions are presented in Section VI.

## II. REVIEW OF THE COVARIANT QUANTIZATION IN THE LORENTZ GAUGE

The need for the longitudinal and scalar photons in a covariant treatment of problems of this kind can be seen by first considering the situation in classical electromagnetism, where the vector potential $\mathbf{A}(\mathbf{r},t)$ and the scalar potential $\Phi(\mathbf{r},t)$ form the components $A^\mu$ of a relativistic four-vector $\mathbf{A}$. Using the Lorentz gauge gives rise to the usual wave equations with retarded solutions, which are manifestly covariant. Quantizing all three components of $\mathbf{A}(\mathbf{r},t)$ as well as the scalar potential $\Phi(\mathbf{r},t)$ in the Lorentz gauge is necessary in order to maintain the manifest covariance of the theory under Lorentz transformations.

The quantization of the electromagnetic field in the Coulomb gauge gives rise to the usual operators $\hat{a}^\dagger_{\perp 1}(\mathbf{k})$ and $\hat{a}^\dagger_{\perp 2}(\mathbf{k})$ that create photons with wave vector $\mathbf{k}$ and transverse polarization $\boldsymbol{\varepsilon}_1$ and $\boldsymbol{\varepsilon}_2$. These photons represent the transverse part of the vector potential $\mathbf{A}(\mathbf{r},t)$, while the scalar potential $\Phi(\mathbf{r},t)$ is not quantized. This is convenient in several respects, but it is not manifestly covariant; under a Lorentz transformation, components of the field that were not quantized in one reference frame will be quantized in another reference frame.

All three components of the vector potential $\mathbf{A}(\mathbf{r},t)$ as well as the scalar potential $\Phi(\mathbf{r},t)$ are quantized in a covariant treatment in the Lorentz gauge [1-6]. This gives rise to a new set of photon creation operators $\hat{a}^\dagger_l(\mathbf{k})$ and $\hat{a}^\dagger_s(\mathbf{k})$ that create photons associated with the longitudinal part of $\mathbf{A}(\mathbf{r},t)$ and the scalar potential $\Phi(\mathbf{r},t)$, respectively. These photons are referred to as longitudinal and scalar (or temporal) photons [1,2].

Physically, a beam of light is polarized only in the transverse direction and the longitudinal and scalar photons must be fictitious in the absence of any charge or current distributions. Gupta [1] and Bleuler [2] independently proposed a solution to this problem, in which negative probabilities are associated with the scalar photons in such a way that they cancel out the effects of the longitudinal photons in a radiation field. The usual inner product $\langle\phi|\psi\rangle$ between two states $|\phi\rangle$ and $|\psi\rangle$ is replaced with an indefinite metric $(\phi|\psi)$, which has the property that states with an odd number of scalar photons have a negative norm. An excellent description of the indefinite metric and the quantization in the Lorentz gauge is given in the text by Cohen-Tannoudji et al. [6] and we will use notation similar to theirs. It should be emphasized that this is the currently accepted covariant formulation of quantum electrodynamics [3-6].

The adjoint of the operators $\hat{a}_{\perp 1}$, $\hat{a}_{\perp 2}$, $\hat{a}_l$, and $\hat{a}_s$ with respect to the new (indefinite) metric will be denoted by $\hat{a}^\mathsf{T}_{\perp 1}$, $\hat{a}^\mathsf{T}_{\perp 2}$, $\hat{a}^\mathsf{T}_l$, and $\hat{a}^\mathsf{T}_s$. The desired properties of the indefinite metric can be obtained if we postulate the following commutation relations:

$$[\hat{a}_{\perp\varepsilon}(\mathbf{k}),\hat{a}^\mathsf{T}_{\perp\varepsilon}(\mathbf{k}')] = \delta_{\mathbf{k}-\mathbf{k}'}$$
$$[\hat{a}_l(\mathbf{k}),\hat{a}^\mathsf{T}_l(\mathbf{k}')] = \delta_{\mathbf{k}-\mathbf{k}'} \qquad (1)$$
$$[\hat{a}_s(\mathbf{k}),\hat{a}^\mathsf{T}_s(\mathbf{k}')] = -\delta_{\mathbf{k}-\mathbf{k}'}.$$

It is important to note the minus sign in the commutation relation for $\hat{a}_s(\mathbf{k})$, which plays an important role in maintaining the consistency of the theory. This is equivalent to taking $\hat{a}^\mathsf{T}_s = -\hat{a}^\dagger_s$, where $\hat{a}^\dagger_s$ is the adjoint of $\hat{a}_s$ with respect to the usual inner product. It is then straightforward to show that states with an odd number of scalar photons have a norm of -1.

The Hamiltonian for the radiation field is chosen to be Hermitian with respect to the indefinite metric and it has the form

$$\hat{H}_R = \int d^3k \frac{\hbar\omega}{2}[(\hat{a}^\mathsf{T}_{\perp 1}\hat{a}_{\perp 1} + \hat{a}_{\perp 1}\hat{a}^\mathsf{T}_{\perp 1}) + (\hat{a}^\mathsf{T}_{\perp 2}\hat{a}_{\perp 2} + \hat{a}_{\perp 2}\hat{a}^\mathsf{T}_{\perp 2}) + (\hat{a}^\mathsf{T}_l\hat{a}_l + \hat{a}_l\hat{a}^\mathsf{T}_l) - (\hat{a}^\mathsf{T}_s\hat{a}_s + \hat{a}_s\hat{a}^\mathsf{T}_s)]. \qquad (2)$$

The minus sign on the last term in Eq. (2) gives a positive energy for the scalar photons when combined with the commutation relations of Eq. (1).

The components of the vector and scalar potential operators are given by

$$\hat{A}_i(\mathbf{r}) = \int d^3\mathbf{k}\sqrt{\frac{\hbar}{2\varepsilon_0\omega(2\pi)^3}}[\hat{a}_i(\mathbf{k})e^{i\mathbf{k}\cdot\mathbf{r}} + \hat{a}^\mathsf{T}_i(\mathbf{k})e^{-i\mathbf{k}\cdot\mathbf{r}}]$$
$$\hat{A}_s(\mathbf{r}) = \int d^3\mathbf{k}\sqrt{\frac{\hbar}{2\varepsilon_0\omega(2\pi)^3}}[\hat{a}_s(\mathbf{k})e^{i\mathbf{k}\cdot\mathbf{r}} + \hat{a}^\mathsf{T}_s(\mathbf{k})e^{-i\mathbf{k}\cdot\mathbf{r}}]. \qquad (3)$$

Here $\varepsilon_0$ is the permittivity of free space and $\hat{A}_s \equiv \hat{\Phi}/c$. $\hat{A}_i(\mathbf{r})$ and $\hat{A}_s(\mathbf{r})$ form the components of a four-vector $\hat{A}^\mu(r)$.

The particle field operators $\hat{\psi}(\mathbf{r},t)$ and $\hat{\psi}^\dagger(\mathbf{r},t)$ are defined as usual in the second-quantized Dirac theory. If we let $\hat{H}_P$ denote the Dirac Hamiltonian describing the particles in the absence of any interaction with the electromagnetic field, then the total Hamiltonian for the system of particles and electromagnetic field has the form [6]

$$\hat{H} = \hat{H}_P + \hat{H}_R + \hat{H}_I. \quad (4)$$

Here the interaction Hamiltonian is given by

$$\hat{H}_I = \int d^3\mathbf{r}\, j_\mu(\mathbf{r}) A^\mu(\mathbf{r}) \\ = \int d^3\mathbf{r} \left[ -\hat{\mathbf{j}}(\mathbf{r}) \cdot \hat{\mathbf{A}}(\mathbf{r}) + c\hat{\rho}(\mathbf{r}) \hat{A}_s(\mathbf{r}) \right] \quad (5)$$

where $\hat{\rho}(\mathbf{r})$ and $\hat{\mathbf{j}}(\mathbf{r})$ are the usual charge and current density operators.

The time dependence of $\hat{\psi}(\mathbf{r},t)$ in the Heisenberg picture can be calculated from

$$\frac{d\hat{\psi}(\mathbf{r},t)}{dt} = \frac{1}{i\hbar}\left[\hat{\psi}(\mathbf{r},t), \hat{H}\right]. \quad (6)$$

Evaluating the commutators gives

$$i\hbar \frac{d\hat{\psi}(\mathbf{r})}{dt} = \left[\beta mc^2 + qc\hat{A}_s(\mathbf{r}) + c\boldsymbol{\alpha}\cdot\left(\frac{\hbar}{i}\nabla - q\hat{\mathbf{A}}(\mathbf{r})\right)\right]\hat{\psi}(\mathbf{r}). \quad (7)$$

This is the usual second-quantized Dirac equation, where $\boldsymbol{\alpha}$ and $\beta$ are the Dirac matrices, which is thus consistent with the Hamiltonian of Eq. (4).

The negative norms that arise from the anomalous commutation relations for $\hat{a}_s(\mathbf{k})$ and $\hat{a}_s^\dagger(\mathbf{k})$ could, in principle, lead to events that occur with negative probabilities. This can be avoided [1-3] if we restrict the physical states $|\chi\rangle$ of the field to those that satisfy the additional (subsidiary) condition

$$\left[\hat{a}_l(\mathbf{k}) - \hat{a}_s(\mathbf{k}) + \hat{\lambda}(\mathbf{k})\right]|\chi\rangle = 0. \quad (8)$$

Here the operator $\hat{\lambda}(\mathbf{k})$ is defined by

$$\hat{\lambda}(\mathbf{k}) = \frac{c}{\omega\sqrt{2\varepsilon_0 \hbar \omega}} \hat{\rho}(\mathbf{k}). \quad (9)$$

It can be shown that the subsidiary condition of Eq. (8) corresponds to the Fourier transform of the Lorentz condition

$$\nabla \cdot \hat{\mathbf{A}} + \frac{1}{c^2}\frac{\partial \hat{\Phi}}{\partial t} = 0. \quad (10)$$

Only the positive frequency components of Eq. (10) are included in the subsidiary condition of Eq. (8), since it would be impossible to satisfy that condition with the negative frequency components included.

In the absence of any charges, $\hat{\lambda}(\mathbf{k}) = 0$ and the subsidiary condition takes on the simpler form

$$\left[\hat{a}_l(\mathbf{k}) - \hat{a}_s(\mathbf{k})\right]|\chi\rangle = 0. \quad (11)$$

It can be seen from Eq. (11) that the probability amplitudes to annihilate a longitudinal or scalar photon must be equal for a pure radiation field, such as a beam of light. The probability of detecting a scalar photon is then equal and opposite to that for detecting a longitudinal photon, and the total detection probability is just that of the transverse photons. This is an example of the way in which the subsidiary condition ensures that no physically observable event can occur with negative probability.

It can be shown from the commutation relations of Eq. (3) that

$$\hat{a}_s \hat{a}_s^\dagger |0\rangle = -|0\rangle \quad (12)$$

where $|0\rangle$ is the vacuum state with no photons. This result is very useful when calculating the relevant matrix elements for use in perturbation theory.

This paper is primarily concerned with a comparison of the entanglement obtained in the Lorentz and Coulomb gauges. The velocities of the particles will be assumed to be much less than the speed of light and only the electromagnetic field will be treated covariantly. In that limit Eq. (4) reduces [26] to the usual nonrelativistic Hamiltonian:

$$\hat{H} = \frac{1}{2m}\left(\frac{\hbar}{i}\nabla - \frac{e}{c}\hat{\mathbf{A}}\right)^2 + q\hat{\Phi} + \hat{H}_R. \quad (13)$$

The entangled states of interest can now be calculated in the Lorentz gauge using the Hamiltonian of Eq. (13) and the commutation relations of Eq. (1). A fully relativistic example including the use of the Dirac theory for the particles will be described elsewhere.

### III. ENTANGLEMENT IN THE LORENTZ GAUGE

Any physical interaction between two systems can generate entanglement between them. In the Lorentz gauge, the exchange of longitudinal and scalar photons can produce an entangled state between the two harmonic oscillators shown in Fig. 1. For example, oscillator A can emit a longitudinal or scalar photon and make a transition from its excited state to its ground state, after which oscillator B can absorb the photon and make a transition from its ground state to its excited state.

For simplicity, we will consider the case of one-dimensional harmonic oscillators in which the motion of the charged particles is confined to lie along the direction between the two oscillators, which we will take to be the $x$ axis. Similar results are expected for two-level atoms, but the use of one-dimensional harmonic oscillators simplifies the matrix elements needed for perturbation theory calculations. This assumption limits the dipole moments of the oscillators to the $\hat{x}$ direction while the wave vectors of the photons can be in any direction.

In the absence of any coupling to the electromagnetic field, the energy eigenstates of the system are product states such as

$$|\psi_0\rangle = |1_A\rangle|0_B\rangle. \quad (14)$$

Here $|1_A\rangle$ denotes the first excited state of oscillator $A$ while $|0_B\rangle$ denotes the ground state of oscillator $B$. The difference in the energies of the first excited states will be denoted by $\delta E = \hbar(\omega_B - \omega_A)$, where $\omega_A$ and $\omega_B$ are the unperturbed resonant frequencies of the two oscillators. It will be assumed that $\delta E \ll \hbar\omega_A$.

If we include the coupling to the electromagnetic field, then the exchange of virtual photons will perturb the eigenstate of Eq. (14) to give a state of the form

$$|\psi\rangle = |1_A\rangle|0_B\rangle + \varepsilon|0_A\rangle|1_B\rangle. \quad (15)$$

Here $\varepsilon$ is a complex probability amplitude whose value can be calculated using perturbation theory. (Eq. (15) will not be normalized in order to simplify the notation.) The fact that $\delta E$ is small causes other possible terms in the perturbed state of Eq. (15), such as those in which both oscillators occupy higher excited states, to be negligible in comparison as will be seen below.

One of the main goals of this paper is to compare the value of $\varepsilon$ as calculated in the Lorentz gauge to the corresponding value in the Coulomb gauge. The exchange of transverse photons will be neglected here since it has the same effect in both gauges. As mentioned above, the velocities of the particles will be assumed to be small compared to the speed of light.

The value of $\varepsilon$ can be calculated to second order in the charge $q$ of the particles using steady-state perturbation theory [26]:

$$|\psi^{(2)}\rangle = \sum_m \sum_l \frac{|m\rangle\langle m|\hat{H}_I|l\rangle\langle l|\hat{H}_I|n\rangle}{(E_n - E_m)(E_n - E_l + i\eta\hbar)}. \quad (16)$$

Here $|\psi^{(2)}\rangle$ is the second-order change in the eigenstate, $|n\rangle$ is the initial state of Eq. (14), and $|l\rangle$ and $|m\rangle$ are complete sets of possible virtual states. The $i\eta\hbar$ term avoids a singularity if $E_n = E_l$, where the limit of $\eta \to 0$ is taken as usual. The matrix elements do not allow transitions to intermediate states where $|l\rangle = |n\rangle$ and the corresponding term can be omitted from the sum over $l$. The value of $\varepsilon$ in Eq. (15) corresponds to the coefficient of the virtual state $|m\rangle = |0_A\rangle|1_B\rangle$.

There are two basic kinds of processes that can produce an entangled state of the form shown in Eq. (15). The most intuitive process is one in which oscillator A emits a longitudinal or scalar photon and makes a transition from its excited state to its ground state, after which oscillator B absorbs the photon and makes a transition from its ground state to its excited state. Diagrams of this kind will be referred to as type I.

In addition to this, it is possible for oscillator B to emit a longitudinal or scalar photon and make a transition from its ground state to its excited state, even though the energy $E_n - E_l$ in the denominator of Eq. (16) is larger in magnitude than for the more intuitive type I processes described above. In that case, oscillator A can subsequently absorb the photon and make a transition from its excited state to its ground state. Counter-intuitive diagrams of this kind will be referred to as type II. Their contribution to $\varepsilon$ will be found to be comparable to that from the more intuitive type I processes because there is less cancellation between the probably amplitudes to exchange longitudinal and scalar photons for a type II process.

We will first consider a type I process in which oscillator A emits a virtual scalar photon that is absorbed by oscillator B. Using Eqs. (3) and (5) gives the matrix element for the emission of the scalar photon

$$\langle l|\hat{H}'_S|n\rangle = \langle 0_A, \mathbf{k}_S|\int d^3\mathbf{r}\, c\hat{\rho}\hat{A}_S|1_A, 0_S\rangle = \\ qc\sqrt{\frac{\hbar}{2\varepsilon_0\omega_\gamma(2\pi)^3}}\int d^3\mathbf{r}\, \psi^*_{0A}(\mathbf{r})e^{-i\mathbf{k}_S\cdot\mathbf{r}}\psi_{1A}(\mathbf{r}). \quad (17)$$

Here $|1_A, 0_S\rangle$ denotes the initial state with oscillator $A$ in its first excited state and no scalar photons, while $\langle 0_A, \mathbf{k}_S|$ denotes the intermediate state with oscillator $A$ in its ground state and a scalar photon with wave vector $\mathbf{k}_S$ and frequency $\omega_\gamma = ck_S$. The interaction Hamiltonian for the scalar photons has been denoted by $\hat{H}'_S$. The nonrelativistic limit of the Hamiltonian, Eq. (13), was used to express the matrix elements in terms of the harmonic oscillator wave functions $\psi_{0A}(\mathbf{r})$ and $\psi_{1A}(\mathbf{r})$.

The matrix element of Eq. (17) could be evaluated in the dipole approximation, but it will be found that that would lead to divergent integrals. The divergence can be eliminated by retaining the exponential factor and not making the dipole approximation, in which case the integral of Eq. (17) can be evaluated to give

$$\langle l|\hat{H}'_S|n\rangle = qc\sqrt{\frac{\hbar}{2\varepsilon_0\omega_\gamma(2\pi)^3}}(-i\mathbf{k}_S\cdot\mathbf{d}) \times e^{-i\mathbf{k}_S\cdot\mathbf{r}_{A0}}e^{-(\mathbf{k}_S\cdot\mathbf{d})^2/2}. \quad (18)$$

Here $\mathbf{r}_{A0}$ is the location of the center of oscillator $A$ and $d$ is the dipole moment of the harmonic oscillator (aside from the charge) given by

$$d = \langle 0_A|\hat{x}|1_A\rangle = \sqrt{\frac{\hbar}{2m\omega_A}} \quad (19)$$

where $m$ is the mass of the particles.

The matrix element for the absorption of the scalar photon by oscillator $B$ can be evaluated in the same way to give

$$\langle m|\hat{H}'_S|l\rangle = -qc\sqrt{\frac{\hbar}{2\varepsilon_0\omega_\gamma(2\pi)^3}}(i\mathbf{k}_S\cdot\mathbf{d})e^{i\mathbf{k}_S\cdot\mathbf{r}_{B0}}e^{-(\mathbf{k}_S\cdot\mathbf{d})^2/2} \quad (20)$$

where it has been assumed that the two oscillators have the same dipole moment. It is important to note the minus sign in front of this equation, which comes from Eq. (12). The contribution from the longitudinal and scalar photons will be found to nearly cancel for a type I process as a result of this minus sign.

Converting the sum of Eq. (16) to an integral and inserting Eqs. (18) and (20) gives the contribution $|\psi^{(2)}\rangle_{IS}$ of this process to the perturbed state as

$$|\psi^{(2)}\rangle_{IS} = \frac{q^2c}{2\varepsilon_0\delta E(2\pi)^3}\int d^3\mathbf{k}_S\frac{(\mathbf{k}_S\cdot\mathbf{d})^2}{k_S} \times \exp[-i\mathbf{k}_S\cdot(\mathbf{r}_{A0}-\mathbf{r}_{B0})]e^{-(\mathbf{k}_S\cdot\mathbf{d})^2}\frac{1}{\omega_A-\omega_\gamma+i\eta}|0_A\rangle|1_B\rangle. \quad (21)$$

The contribution $|\psi^{(2)}\rangle_{IIS}$ from the emission of a scalar photon by oscillator $B$ and its absorption by oscillator $A$ can be calculated in a similar way. The main difference is the energy of the intermediate state, with the result that

$$|\psi^{(2)}\rangle_{IIS} = \frac{q^2c}{2\varepsilon_0\delta E(2\pi)^3}\int d^3\mathbf{k}_S\frac{(\mathbf{k}_S\cdot\mathbf{d})^2}{k_S} \times \exp[i\mathbf{k}_S\cdot(\mathbf{r}_{A0}-\mathbf{r}_{B0})]e^{-(\mathbf{k}_S\cdot\mathbf{d})^2}\frac{1}{-\omega_B-\omega_\gamma+i\eta}|0_A\rangle|1_B\rangle. \quad (22)$$

We now calculate the contribution from a type I process in which oscillator A emits a longitudinal photon that is absorbed by oscillator B. The matrix elements for the emission of a longitudinal photon by oscillator A involve $-\hat{\mathbf{j}}(\mathbf{r})\cdot\mathbf{A}_l(\mathbf{r})$, where $\mathbf{A}_l(\mathbf{r})$ is the longitudinal part of the vector potential. In the nonrelativistic limit this gives

$$\langle l|\hat{H}'_l|n\rangle = -\frac{q}{2m}\sqrt{\frac{\hbar}{2\varepsilon_0\omega_\gamma(2\pi)^3}} \times \int d^3\mathbf{r}\psi^*_{0A}(\mathbf{r})(\hat{\mathbf{k}}_l\cdot\hat{\mathbf{d}})\left(e^{-i\mathbf{k}_l\cdot\mathbf{r}}\frac{\hbar}{i}\frac{\partial}{\partial x}+\frac{\hbar}{i}\frac{\partial}{\partial x}e^{-i\mathbf{k}_l\cdot\mathbf{r}}\right)\psi_{1A}(\mathbf{r}). \quad (23)$$

Here $\omega_\gamma = ck_l$, $\hat{H}'_l$ denotes the interaction Hamiltonian associated with longitudinal photons with wave vector $\mathbf{k}_l$, while $\hat{\mathbf{k}}_l$ and $\hat{\mathbf{d}}$ denote the corresponding unit vectors. Evaluating this integral gives

$$\langle l|\hat{H}'_l|n\rangle = -qc\frac{\omega_A}{\omega_\gamma}\sqrt{\frac{\hbar}{2\varepsilon_0\omega_\gamma(2\pi)^3}} \times (-i\mathbf{k}_l\cdot\mathbf{d})e^{-i\mathbf{k}_l\cdot\mathbf{r}_{A0}}e^{-(\mathbf{k}_l\cdot\mathbf{d})^2/2}. \quad (24)$$

This matrix element for the emission of a longitudinal photon differs from that for the emission of a scalar photon, Eq. (18), by a minus sign and a factor of $\omega_A/\omega_\gamma$. This ensures that the subsidiary condition of Eq. (11) would be satisfied exactly for the emission of real longitudinal and scalar photons with $\omega_A = \omega_\gamma$, for example.

The matrix element for the absorption of a longitudinal photon by oscillator B can be evaluated in the same way to give

$$\langle m|\hat{H}'_l|l\rangle = -qc\frac{\omega_B}{\omega_\gamma}\sqrt{\frac{\hbar}{2\varepsilon_0\omega_\gamma(2\pi)^3}}(i\mathbf{k}_l\cdot\mathbf{d})e^{i\mathbf{k}_l\cdot\mathbf{r}_{B0}}e^{-(\mathbf{k}_l\cdot\mathbf{d})^2/2}. \quad (25)$$

The contribution $|\psi^{(2)}\rangle_{Il}$ of this process to the perturbed eigenstate is then given by combining Eqs. (16), (24), and (25) to obtain

$$|\psi^{(2)}\rangle_{II} = -\frac{q^2 c}{2\varepsilon_0 \delta E (2\pi)^3} \int d^3 \mathbf{k}_l \frac{(\mathbf{k}_l \cdot \mathbf{d})^2}{k_l}$$
$$\times \exp[-i\mathbf{k}_l \cdot (\mathbf{r}_{A0} - \mathbf{r}_{B0})] e^{-(\mathbf{k}_l \cdot \mathbf{d})^2} \frac{\omega_A \omega_B}{\omega_\gamma^2} \frac{1}{\omega_A - \omega_\gamma + i\eta} |0_A\rangle |1_B\rangle. \quad (26)$$

Similarly, the contribution $|\psi^{(2)}\rangle_{III}$ from the emission of a longitudinal photon by oscillator B and its absorption by oscillator A can be shown to be

$$|\psi^{(2)}\rangle_{III} = -\frac{q^2 c}{2\varepsilon_0 \delta E (2\pi)^3} \int d^3 \mathbf{k}_l \frac{(\mathbf{k}_l \cdot \mathbf{d})^2}{k_l}$$
$$\times \exp[i\mathbf{k}_l \cdot (\mathbf{r}_{A0} - \mathbf{r}_{B0})] e^{-(\mathbf{k}_l \cdot \mathbf{d})^2} \frac{\omega_A \omega_B}{\omega_\gamma^2} \frac{1}{-\omega_B - \omega_\gamma + i\eta} |0_A\rangle |1_B\rangle. \quad (27)$$

The nonrelativistic Hamiltonian of Eq. (13) also contains a term proportional to $q^2 \mathbf{A}^2 / 2mc^2$, which can simultaneously emit or absorb two longitudinal photons. The two photons are emitted or absorbed at the same location, however. As a result, the $\mathbf{A}^2$ term cannot contribute in order $q^2$ to the effects of interest here, which require a transition in the state of both oscillators.

Combining Eqs. (21), (22), (26), and (27) gives the total second-order contribution to the entangled state. This corresponds to a value of $\varepsilon$ given by

$$\varepsilon = -\frac{q^2}{\varepsilon_0 \delta E (2\pi)^3} \int d^3 \mathbf{k} \frac{(\mathbf{k} \cdot \mathbf{d})^2}{k^2} \cos[\mathbf{k} \cdot (\mathbf{r}_{A0} - \mathbf{r}_{B0})]$$
$$\times e^{-(\mathbf{k} \cdot \mathbf{d})^2} \left\{ \frac{1}{2}\left[\frac{(\omega_A \omega_B - \omega_\gamma^2)}{\omega_\gamma}\right]\left[\frac{1}{\omega_A - \omega_\gamma + i\eta} - \frac{1}{\omega_B + \omega_\gamma}\right] \right\} \quad (28)$$

where here $\omega_\gamma = ck$. The integral can be evaluated if we expand the quantity inside the curly brackets in a power series in $\delta E$, which has been assumed to be small. This gives

$$\varepsilon = -\frac{q^2}{\varepsilon_0 \delta E (2\pi)^3} \int d^3 \mathbf{k} \frac{(\mathbf{k} \cdot \mathbf{d})^2}{k^2} \cos[\mathbf{k} \cdot (\mathbf{r}_{A0} - \mathbf{r}_{B0})]$$
$$\times e^{-(\mathbf{k} \cdot \mathbf{d})^2} \left\{ 1 + \frac{1}{2} \frac{(\omega_A^2 + \omega_\gamma^2)}{\omega_\gamma (\omega_A + \omega_\gamma)(\omega_A - \omega_\gamma + i\eta)}\left(\frac{\delta E}{\hbar}\right) \right.$$
$$\left. + \frac{1}{2} \frac{1}{(\omega_A + \omega_\gamma)^2}\left(\frac{\delta E}{\hbar}\right)^2 + ... \right\}. \quad (29)$$

The integrals in Eq. (29) can be evaluated using contour integration and other techniques in the limit where the dipole moment is much less than the distance between the oscillators, or $d \ll L$. In that limit, the value of the coefficient $\varepsilon$ in the entangled state of Eq. (15) reduces to

$$\varepsilon_L = \frac{d^2 q^2}{2\pi \varepsilon_0 \delta E L^3}\left[1 - \frac{1}{2\pi}\left(\frac{\delta E}{\hbar \omega_L}\right) + \frac{1}{2}\left(\frac{\delta E}{\hbar \omega_A}\right)^2 + ...\right] \quad (30)$$

where $\omega_L \equiv c/L$. Eq. (30) corresponds to the principal value integral of Eq. (29). This result has been labeled with a subscript $L$ to indicate that it was calculated in the Lorentz gauge. The corresponding value in the Coulomb gauge is calculated in the next section. Neither calculation includes the contribution from the exchange of transverse photons, which is the same in both gauges and of no interest here.

## IV. ENTANGLEMENT IN THE COULOMB GAUGE

It was shown in the previous section that the exchange of longitudinal and scalar photons can produce an entangled state of two harmonic oscillators. The corresponding calculation will now be performed in the Coulomb gauge where the Coulomb potential is not quantized and there is no longitudinal component of the vector potential.

The total Hamiltonian of the system is still given by Eq. (4) where $\hat{H}_P$ is the same as before. But now the Hamiltonian for the radiation field in the absence of interaction is given by

$$\hat{H}_R = \int d^3 k \frac{\hbar \omega}{2}[(\hat{a}_{\perp 1}^\dagger \hat{a}_{\perp 1} + \hat{a}_{\perp 1} \hat{a}_{\perp 1}^\dagger) + (\hat{a}_{\perp 2}^\dagger \hat{a}_{\perp 2} + \hat{a}_{\perp 2} \hat{a}_{\perp 2}^\dagger)] \quad (31)$$

which only includes the energies of the transverse photons. The interaction Hamiltonian becomes [6]

$$\hat{H}_I = -\int d^3 \mathbf{r} \hat{\mathbf{j}}(\mathbf{r}) \cdot \hat{\mathbf{A}}_\perp(\mathbf{r}) + \frac{1}{8\pi \varepsilon_0} \iint d^3 \mathbf{r} d^3 \mathbf{r}' \frac{\hat{\rho}(\mathbf{r})\hat{\rho}(\mathbf{r}')}{|\mathbf{r} - \mathbf{r}'|}. \quad (32)$$

Now $\hat{H}_I$ only involves the transverse component $\hat{\mathbf{A}}_\perp(\mathbf{r})$ of the vector potential, while the second term in Eq. (32) corresponds to the classical Coulomb energy of a charge distribution; the operators $\hat{A}_l$ and $\hat{\Phi}$ have been eliminated.

The comparison with the Lorentz gauge is more apparent if we use the Fourier transform of the Coulomb interaction:

$$\frac{1}{8\pi\varepsilon_0}\iint d^3\mathbf{r}\,d^3\mathbf{r}'\frac{\hat{\rho}(\mathbf{r})\hat{\rho}(\mathbf{r}')}{|\mathbf{r}-\mathbf{r}'|} = \int d^3\mathbf{k}\frac{\hat{\rho}^\dagger(\mathbf{k})\hat{\rho}(\mathbf{k})}{2\varepsilon_0 k^2}. \quad (33)$$

Since the interaction Hamiltonian is already second order in $q$, we only need to use first-order perturbation theory here. The relevant perturbation to the state vector is now given by

$$|\psi^{(2)}\rangle = \sum_m |m\rangle \frac{\langle m|\hat{H}'_C|n\rangle}{E_n - E_m} \quad (34)$$

where we have denoted the Coulomb interaction Hamiltonian of Eq. (33) by $\hat{H}'_C$. The effects of the transverse photons will be neglected once again.

The operator $\hat{\rho}(\mathbf{k})$ can produce a transition of one of the harmonic oscillators from one state to another, while $\hat{\rho}^\dagger(\mathbf{k})\hat{\rho}(\mathbf{k})$ can produce simultaneous transitions in both oscillators. The matrix elements of $\hat{\rho}(\mathbf{k})$ can be shown to be the same as those of Eqs. (17) and (20) aside from a constant, where the exponential factor now comes from the Fourier transform.

There are four different ways in which the virtual state $|0_A\rangle|1_B\rangle$ can be produced starting from $|1_A\rangle|0_B\rangle$. For example, there is a contribution $|\psi^{(2)}\rangle_i$ in which operator $\hat{\rho}(\mathbf{k})$ produces a transition of oscillator $A$ from its excited state to its ground state while the operator $\hat{\rho}^\dagger(\mathbf{k}) = \hat{\rho}(-\mathbf{k})$ simultaneously produces a transition of oscillator $B$ from its ground state to its excited state. From Eq. (33), the matrix element $M_i$ associated with this process is given by

$$M_i = \int d^3\mathbf{k}\frac{\langle 1_B|\hat{\rho}(-\mathbf{k})|0_B\rangle\langle 0_A|\hat{\rho}(\mathbf{k})|1_A\rangle}{2\varepsilon_0 k^2}. \quad (35)$$

It is also possible for operator $\hat{\rho}(-\mathbf{k})$ to produce a transition of oscillator $A$ from its excited state to its ground state, while the operator $\hat{\rho}(\mathbf{k})$ produces a transition of oscillator $B$ from its ground state to its excited state. The corresponding matrix element is given by

$$M_{ii} = \int d^3\mathbf{k}\frac{\langle 1_B|\hat{\rho}(\mathbf{k})|0_B\rangle\langle 0_A|\hat{\rho}(-\mathbf{k})|1_A\rangle}{2\varepsilon_0 k^2}. \quad (36)$$

In principle, it is also possible for the operator $\hat{\rho}(\mathbf{k})$ to annihilate a particle from oscillator $A$ in the state $|1_A\rangle$ and recreate it in oscillator $B$ in the state $|1_B\rangle$, with a similar effect from operator $\hat{\rho}(-\mathbf{k})$. This corresponds to matrix elements of the form

$$M_{iii} = \int d^3\mathbf{k}\frac{\langle 0_A|\hat{\rho}(-\mathbf{k})|0_B\rangle\langle 1_B|\hat{\rho}(\mathbf{k})|1_A\rangle}{2\varepsilon_0 k^2}. \quad (37)$$

Matrix elements of this kind are negligibly small in the limit of $d \ll L$, since the overlap of the wave functions of the two oscillators decreases exponentially with their separation. We therefore neglect $M_{iii}$ and the corresponding matrix element $M_{iv}$ with $\hat{\rho}(\mathbf{k})$ and $\hat{\rho}(-\mathbf{k})$ interchanged.

Inserting the values of $M_i$ and $M_{ii}$ into Eq. (34) gives

$$|\psi^{(2)}\rangle = -\frac{q^2}{\varepsilon_0 \delta E (2\pi)^3}\int d^3\mathbf{k}\frac{(\mathbf{k}\cdot\mathbf{d})^2}{k^2}$$
$$\times \cos[\mathbf{k}_l\cdot(\mathbf{r}_{A0}-\mathbf{r}_{B0})]e^{-(\mathbf{k}\cdot\mathbf{d})^2}|0_A\rangle|1_B\rangle. \quad (38)$$

Evaluating this integral as before gives the result that

$$\varepsilon_C = \frac{d^2 q^2}{2\pi\varepsilon_0 \delta E L^3}. \quad (39)$$

A subscript C has been added to indicate that this is the coefficient of the term $|0_A\rangle|1_B\rangle$ in the entangled state as calculated in the Coulomb gauge.

## V. COMPARISON OF THE RESULTS IN THE TWO GAUGES

A comparison of $\varepsilon_C$ from Eq. (39) with $\varepsilon_L$ from Eq. (30) shows that the exchange of longitudinal and scalar photons in the Lorentz gauge gives an entangled state that is very different in form from that obtained in the Coulomb gauge. The leading term in the expansion of Eq. (30) is the same as that in Eq. (39), but the Coulomb gauge does not have the same dependence on the energy difference $\delta E$ as is obtained in the Lorentz gauge. This can be understood in part as being due to the presence of the energies of the intermediate states in the denominators of the second-order perturbation theory of Eq. (16), which does not occur in the Coulomb gauge treatment. In addition, the matrix elements associated with the longitudinal photons in Eq. (25) do not have any direct counterpart in the Coulomb gauge. Thus it is not surprising that the form of the entangled state is different in the two cases.

Nevertheless, one would expect the two results to be physically equivalent based on gauge invariance. For classical fields, the Coulomb and Lorentz gauges are related by a gauge transformation of the form

$$\mathbf{A}'(\mathbf{r},t) = \mathbf{A}(\mathbf{r},t) + \nabla\Lambda(\mathbf{r},t)$$

$$\phi'(\mathbf{r},t) = \phi(\mathbf{r},t) - \frac{\partial\Lambda(\mathbf{r},t)}{\partial t} \tag{40}$$

where $\Lambda(\mathbf{r},t)$ is an arbitrary function of position and time [27]. Under such a gauge transformation, the new wave function becomes [26]

$$\psi'(\mathbf{r},t) = e^{-iq\Lambda(\mathbf{r},t)/\hbar}\psi(\mathbf{r},t). \tag{41}$$

The system is physically equivalent in either gauge because $\psi'(\mathbf{r},t)$ is not directly observable and $\rho = \psi^*\psi$ is unchanged by the transformation.

The situation is more complicated for quantized fields in part because the Hilbert spaces have different dimensions. It can be shown [6,28] that the state vector $|\psi_C\rangle$ in the Coulomb gauge should be related to the state vector $|\psi_L\rangle$ in the Lorentz gauge by

$$|\psi_C\rangle = \hat{T}|\psi_L\rangle. \tag{42}$$

The transformation $\hat{T}$ is given by

$$\hat{T} = e^{-ic\int \hat{\rho}(\mathbf{r})\hat{S}(\mathbf{r})d^3r/\hbar} \tag{43}$$

while the operator $\hat{S}(\mathbf{r})$ is defined by

$$\hat{S}(r) = \int d^3\mathbf{k}\sqrt{\frac{\hbar}{2\varepsilon_0\omega(2\pi)^3}}\left[\frac{\hat{a}_s(\mathbf{k})}{i\omega}e^{i\mathbf{k}\cdot\mathbf{r}} - \frac{\hat{a}_s^\mathrm{T}(\mathbf{k})}{i\omega}e^{-i\mathbf{k}\cdot\mathbf{r}}\right]. \tag{44}$$

The operators $\hat{T}$ and $\hat{S}(\mathbf{r})$ are defined in the Schrodinger picture, which will be used throughout this section.

This transformation has the property that the Dirac field operator becomes

$$\hat{T}\hat{\psi}(\mathbf{r})\hat{T}^{-1} = e^{iqc\hat{S}(\mathbf{r})/\hbar}\hat{\psi}(\mathbf{r}), \tag{45}$$

which leaves the charge density unaltered:

$$\hat{T}\hat{\rho}(\mathbf{r})\hat{T}^{-1} = \hat{\rho}(\mathbf{r}). \tag{46}$$

Thus the probability of detecting a particle at any given location is unaffected by such a transformation and the results of the two gauges should be physically equivalent [6].

It will now be shown that the entangled states calculated in the Lorentz and Coulomb gauges are indeed related to each other by the transformation of Eq. (42). Consider the state vector $|\psi_L'\rangle$ obtained by transforming the perturbed state vector calculated in the Coulomb gauge back into the Lorentz gauge:

$$|\psi_L'\rangle \equiv \hat{T}^{-1}|\psi_C\rangle. \tag{47}$$

Here $|\psi_C\rangle$ is the entangled state of Eq. (15) using the value of $\varepsilon_C$ calculated from the Coulomb gauge in Eq. (39). To second order in $q$, the value of $|\psi_L'\rangle$ can be obtained by expanding the transformation $\hat{T}^{-1}$ to second order in $q$ and multiplying by the appropriate term in the expansion of $|\psi_C\rangle$. This gives

$$|\psi_L'\rangle^{(2)} = (\hat{T}^{-1})^{(2)}|\psi_C^{(0)}\rangle + (\hat{T}^{-1})^{(1)}|\psi_C^{(1)}\rangle + (\hat{T}^{-1})^{(0)}|\psi_C^{(2)}\rangle. \tag{48}$$

The superscripts in parentheses correspond to the order of that term in $q$. As shown in the Appendix, the result is that

$$|\psi_L'\rangle = |1_A\rangle|0_B\rangle + \frac{d^2q^2}{2\pi\varepsilon_0\delta EL^3}$$
$$\times\left[1 - \frac{1}{2\pi}\left(\frac{\delta E}{\hbar\omega_L}\right) + \frac{1}{2}\left(\frac{\delta E}{\hbar\omega_A}\right)^2 + ...\right]|0_A\rangle|1_B\rangle. \tag{49}$$

A comparison of Eqs. (49) and (30) shows that the results from the Lorentz and Coulomb gauges are indeed related by the transformation $\hat{T}$ as expected, at least to second order in $q$. The derivation of Eq. (49) assumed that $\delta E \ll \hbar\omega_A$ and that $d \ll L$ as before. It may be worth noting that the two gauges give equivalent results for the contribution from each Fourier component $\mathbf{k}$ of the field individually before any integration is performed. These results show that the entangled states in the Lorentz and Coulomb gauges are physically equivalent despite the difference in their forms.

This result is in agreement with a general proof [6,28] that the Coulomb and Lorentz gauges must give equivalent results. The proof is based on the use of the transformation $\hat{T}$ to transform the Hamiltonian in the Lorentz gauge into the Coulomb gauge. The result is that

$$\hat{T}\hat{H}_L T^{-1} = \hat{H}_P + \hat{H}_R - \int d^3\mathbf{r}\,\hat{\mathbf{j}}(\mathbf{r})\cdot\hat{\mathbf{A}}_\perp(\mathbf{r})$$
$$+ \frac{1}{8\pi\varepsilon_0}\iint d^3r d^3r'\frac{\hat{\rho}(\mathbf{r})\hat{\rho}(\mathbf{r}')}{|\mathbf{r}-\mathbf{r}'|} \tag{50}$$
$$- qc\int d^3\mathbf{r}\,\hat{\psi}^\dagger(\mathbf{r})\alpha\cdot\left[\hat{A}_l(\mathbf{r}) - c\nabla\hat{S}(\mathbf{r})\right]\hat{\psi}(\mathbf{r}).$$

Here $\hat{H}_L$ denotes the Hamiltonian in the Lorentz gauge, as given by Eqs. (2) through (5), while $\hat{A}_l(\mathbf{r})$ is the longitudinal part of the vector potential operator. The next to last term in Eq. (50) corresponds to the usual Coulomb potential, as in Eq. (32). The Fourier transform of the last term in Eq. (50) can be shown to contain only the current operator multiplied by $[\hat{a}_l^\mathsf{T}(\mathbf{k}) - \hat{a}_s^\mathsf{T}(\mathbf{k})]$ and its adjoint. As a result, the longitudinal and scalar photons are generated with equal but opposite probability amplitudes, as in Eq. (11), and their effects cancel out just as they would in the absence of any interaction. Thus the last term in Eq. (50) has no physical effects and can be ignored [6], which leaves us with the Hamiltonian $\hat{H}_C$ in the Coulomb gauge.

The results presented here provide an explicit example of the way in which the Coulomb and Lorentz gauges give equivalent results as would be expected from the proof outlined above. Aside from providing a covariant formulation for quantum information protocols, investigations of this kind are also relevant because any proof could conceivably contain hidden assumptions or other weaknesses. For example, consider classical charge and current distributions $\rho_c(\mathbf{r},t)$ and $\mathbf{j}_c(\mathbf{r},t)$ that are explicit functions of time. How would the transformation $\hat{T}$ work in that case? We start with the Schrodinger equation in the Lorentz gauge

$$i\hbar \frac{d|\psi_L\rangle}{dt} = \hat{H}_L |\psi_L\rangle. \tag{51}$$

Multiplying both sides of the equation by $\hat{T}$ and inserting $\hat{T}^{-1}\hat{T} = \hat{I}$ on the right gives

$$i\hbar \hat{T} \frac{d|\psi_L\rangle}{dt} = \hat{T}\hat{H}_L \hat{T}^{-1} \hat{T}|\psi_L\rangle = \hat{H}_C |\psi_C\rangle. \tag{52}$$

The point is that

$$\hat{T} \frac{d|\psi_L\rangle}{dt} \neq \frac{d}{dt}\left(\hat{T}|\psi_L\rangle\right) \tag{53}$$

if $\hat{T}$ is time dependent, which it is from its definition in Eq. (43) and the fact that $\rho_c(\mathbf{r},t)$ is time dependent. With that in mind, the left hand side of Eq. (52) can be rewritten as

$$i\hbar \hat{T} \frac{d|\psi_L\rangle}{dt} = i\hbar \frac{d(\hat{T}|\psi_L\rangle)}{dt} - i\hbar \frac{d\hat{T}}{dt}|\psi_L\rangle. \tag{54}$$

Combining Eqs. (51) through (54) gives

$$i\hbar \frac{d|\psi_C\rangle}{dt} = \hat{H}_C |\psi_C\rangle + i\hbar \frac{d\hat{T}}{dt}|\psi_L\rangle. \tag{55}$$

It can be seen from Eq. (55) that the transformation $\hat{T}$ does not give the correct Schrodinger equation in the Coulomb gauge if $\hat{T}$ is time dependent. Thus the standard proof is not valid for time-dependent classical charge and current distributions in its current form. The proof outlined above can presumably be generalized to deal with time-dependent classical sources, but this illustrates the importance of considering simple examples, such as the entangled states discussed above.

## VI. SUMMARY AND CONCLUSIONS

The covariant quantization of the electromagnetic field in the Lorentz gauge introduces longitudinal and scalar photons in addition to the transverse photons familiar in the Coulomb gauge [1-6]. For a freely-propagating beam of light, the effects of the longitudinal and scalar photons cancel out and they can be ignored. That is not the case in the presence of charge or current distributions, where the longitudinal and scalar photons can produce observable effects.

It has been shown here that the exchange of longitudinal and scalar photons can produce an entangled eigenstate of two harmonic oscillators that is quite different from that produced in the Coulomb gauge where the longitudinal and scalar photons do not exist. This difference can be understood from the presence of the energies of the intermediate states in the denominators of second-order perturbation theory as well as the different form of the matrix elements.

It was also shown that the entangled state in the Coulomb gauge can be related to that in the Lorentz gauge by a transformation $\hat{T}$ involving the charge density and an operator associated with the scalar photons [6,28]. This transformation leaves the charge density unaltered, as would a gauge transformation for classical fields, and the entangled states in the two gauges are thus physically equivalent. The calculations described here were limited to second order in perturbation theory and it was assumed that $\delta E \ll \hbar\omega_A$ and $d \ll L$.

This example illustrates the importance of carefully considering what is actually observable in an entangled state. The change in the probability amplitude $\varepsilon$ for the $|0_A\rangle|1_B\rangle$ component in the entangled state may seem to suggest that the exchange of longitudinal and scalar photons has produced an additional source of entanglement. One might suppose that we could measure the states occupied by the harmonic oscillators in the unperturbed basis and thus determine the magnitude of $\varepsilon$ using an ensemble of such states. But the state vector is not physically observable and neither are the coefficients in an expansion of the state vector in a particular basis, as this example illustrates. This is also the case for gauge transformations in elementary quantum mechanics using classical fields.

Determining the amount of entanglement present in a quantum system is an ongoing field of investigation [29],

and several different entanglement measures have been introduced, such as the concurrence or entanglement of formation [30]. If we were to simply apply one of these entanglement measures to the state of Eq. (15) in the usual way, we would find that the results depend on the value of the parameter $\varepsilon$. This may seem to indicate that there is a different amount of entanglement in the Coulomb and Lorentz gauges, even though they are physically equivalent. This difficulty may be due in part to the fact that Eq. (15) ignores the virtual photons that are also present in the system in addition to the amplitudes of the oscillator states, and that may have to be taken into account in calculating the total amount of entanglement in the system. These issues are beyond the intended scope of this paper but they illustrate the need for further work in this area.

In view of the many nonclassical effects that arise from quantizing the field, it seems remarkable that it should make no difference whether or not we quantize two of the four components of the field. Simple examples of this kind provide physical insight into the way in which the two gauges are equivalent, which is of fundamental importance. In addition, a manifestly covariant description of entanglement is desirable for a fundamental understanding of experiments based on Bell's inequality, especially in view of the nonlocal collapse of the wave function. Techniques of this kind may also be useful for a manifestly covariant description of the time evolution of systems used for quantum information processing and quantum communications, which may be of practical importance for satellite systems where relativistic effects may become significant.

## ACKNOWLEDGEMENTS

I would like to acknowledge stimulating discussions with G. Gilbert.

## APPENDIX

As discussed in the main text, the Coulomb and Lorentz gauges should be related by the transformation $\hat{T}$ defined in Eq. (43). In this Appendix, the expansion of Eq. (48) will be used to show that the two gauges are equivalent for the situation illustrated in Fig. 1, at least to second order in $q$.

The last term in the expansion of Eq. (48) can be found by noting that the zero-order term in $\hat{T}^{-1}$ is just the identity $\hat{I}$. The second-order term in the state vector in the Coulomb gauge is already given in Eq. (38), so that

$$(\hat{T}^{-1})^{(0)}|\psi^{(2)}\rangle = -\frac{q^2}{\varepsilon_0 \delta E (2\pi)^3} \int d^3\mathbf{k} \frac{(\mathbf{k}\cdot\mathbf{d})^2}{k^2} \quad (A1)$$
$$\times \cos[\mathbf{k}_l \cdot (\mathbf{r}_{A0} - \mathbf{r}_{B0})] e^{-(\mathbf{k}\cdot\mathbf{d})^2} |0_A\rangle|1_B\rangle.$$

The first-order term in $\hat{T}^{-1}$ can be found by expanding the exponential in the definition of $\hat{T}$ in a Taylor series to obtain

$$(\hat{T}^{-1})^{(1)} = ic\int \hat{\rho}(\mathbf{r})\hat{S}(\mathbf{r}) d^3\mathbf{r}/\hbar. \quad (A2)$$

where the operator $\hat{S}$ is defined in Eq. (44). This gives

$$(\hat{T}^{-1})^{(1)} = \frac{ic}{\hbar}\int d^3\mathbf{k}\sqrt{\frac{\hbar}{2\varepsilon_0\omega_k}} \quad (A3)$$
$$\times \left[\frac{1}{i\omega_k}\hat{\rho}(-\mathbf{k})\hat{a}_s(\mathbf{k}) - \frac{1}{i\omega_k}\hat{\rho}(\mathbf{k})\hat{a}_s^\dagger(\mathbf{k})\right].$$

Inserting the matrix elements of $\rho(\mathbf{k})$ from Eqs. (18) and (20) gives

$$(\hat{T}^{-1})^{(1)} = \frac{iqd}{\hbar(2\pi)^{3/2}}\int d^3\mathbf{k}\sqrt{\frac{\hbar}{2\varepsilon_0\omega_k}} e^{-(\mathbf{k}\cdot\mathbf{d})^2/2} \quad (A4)$$
$$\times \hat{\mathbf{k}}\cdot\hat{\mathbf{d}}\sum_i \left[\hat{a}_s(\mathbf{k})e^{i\mathbf{k}\cdot\mathbf{r}_{0i}} + \hat{a}_s^T(\mathbf{k})e^{-i\mathbf{k}\cdot\mathbf{r}_{0i}}\right](\hat{b}_i + \hat{b}_i^\dagger).$$

Here $\hat{b}_i$ and $\hat{b}_i^\dagger$ are the usual raising and lowering operators for the harmonic oscillators located at $\mathbf{r}_{0i}$.

The first-order term in the state vector can be found by rewriting [6] the last term in the Hamiltonian of Eq. (50) as

$$\hat{H}_{ls} = -\int d^3\mathbf{k}\sqrt{\frac{\hbar}{2\varepsilon_0\omega_k}}\{j_l(-\mathbf{k})[a_l(\mathbf{k}) - a_s(\mathbf{k})] \quad (A5)$$
$$+ j_l(\mathbf{k})[a_l^T(\mathbf{k}) - a_s^T(\mathbf{k})]\}.$$

Using first-order perturbation theory and inserting the matrix elements for $j_l(\mathbf{k})$ and $j_l(-\mathbf{k})$ gives

$$|\psi^{(1)}\rangle = \int d^3\mathbf{k}\sqrt{\frac{\hbar}{2\varepsilon_0\omega_k}} e^{-(\mathbf{k}\cdot\mathbf{d})^2/2}\{\frac{i\omega_A dq}{\hbar\omega_A - \hbar\omega_k + i\eta}e^{-i\mathbf{k}\cdot\mathbf{r}_{A0}}|0_A\rangle|0_B\rangle$$
$$+\frac{i\omega_B dq}{\hbar\omega_k + \hbar\omega_B}e^{-i\mathbf{k}\cdot\mathbf{r}_{B0}}|1_A\rangle|1_B\rangle\}\hat{\mathbf{k}}\cdot\hat{\mathbf{d}}[a_l^T(\mathbf{k}) - a_s^T(\mathbf{k})]|0_F\rangle.$$
$$(A6)$$

Combining Eqs. (A4) and (A6) and making use of the commutation relations of Eq. (1) gives

$$(\hat{T}^{-1})^{(1)}|\psi^{(1)}\rangle = -\frac{q^2 d^2}{(2\pi)^3}\int d^3\mathbf{k}\frac{1}{2\varepsilon_0 \omega_k}e^{-(\mathbf{k}\cdot\mathbf{d})^2}(\hat{\mathbf{k}}\cdot\hat{\mathbf{d}})^2$$

$$\times\left\{\frac{\omega_A}{\hbar\omega_A - \hbar\omega_k + i\eta}e^{-i\mathbf{k}\cdot(\mathbf{r}_{A0}-\mathbf{r}_{B0})}\right.$$

$$\left.+\frac{\omega_B}{\hbar\omega_B + \hbar\omega_k}e^{i\mathbf{k}\cdot(\mathbf{r}_{A0}-\mathbf{r}_{B0})}\right\}|0_A\rangle|1_B\rangle|0_F\rangle. \quad (A7)$$

The exponential factors in Eq. (A7) can be rewritten as

$$e^{\pm i\mathbf{k}\cdot(\mathbf{r}_{A0}-\mathbf{r}_{B0})} = \cos[\mathbf{k}\cdot(\mathbf{r}_{A0}-\mathbf{r}_{B0})] \pm i\sin[\mathbf{k}\cdot(\mathbf{r}_{A0}-\mathbf{r}_{B0})]. \quad (A8)$$

The sine term is an odd function of $\mathbf{k}$ and it does not contribute to the integral, so that Eq. (A7) reduces to

$$(\hat{T}^{-1})^{(1)}|\psi^{(1)}\rangle = -\frac{1}{(2\pi)^3}\frac{q^2 d^2}{2\varepsilon_0 \hbar}\int\frac{d^3\mathbf{k}}{\omega_k}e^{-(\mathbf{k}\cdot\mathbf{d})^2}(\hat{\mathbf{k}}\cdot\hat{\mathbf{d}})^2$$

$$\times\left\{\frac{\omega_A}{\omega_A - \omega_k + i\eta\hbar}+\frac{\omega_B}{\omega_B + \omega_k}\right\} \quad (A9)$$

$$\times\cos[\mathbf{k}\cdot(\mathbf{r}_{A0}-\mathbf{r}_{B0})]|0_A\rangle|1_B\rangle|0_F\rangle.$$

The remaining term in Eq. (48) can be found by expanding $\hat{T}^{-1}$ to second order in a Taylor series expansion:

$$(\hat{T}^{-1})^{(2)} = -\frac{c^2}{2\hbar^2}\int\hat{\rho}(\mathbf{r})\hat{S}(\mathbf{r})d^3\mathbf{r}\int\hat{\rho}(\mathbf{r}')\hat{S}(\mathbf{r}')d^3\mathbf{r}'. \quad (A10)$$

Inserting the matrix elements for $\rho(\mathbf{k})$ and using the commutation relations as before gives

$$(\hat{T}^{-1})^{(2)}|\psi^{(0)}\rangle = \frac{1}{(2\pi)^3}\frac{q^2 d^2}{2\varepsilon_0}\int\frac{d^3\mathbf{k}}{\hbar\omega_k}e^{-(\mathbf{k}\cdot\mathbf{d})^2}(\hat{\mathbf{k}}\cdot\hat{\mathbf{d}})^2 \quad (A11)$$

$$\times\cos[\mathbf{k}\cdot(\mathbf{r}_{A0}-\mathbf{r}_{B0})]|0_A\rangle|1_B\rangle|0_F\rangle$$

where we have used the form of $|\psi^{(0)}\rangle$ from Eq. (14). Virtual states containing more than one photon have been neglected as in the text.

Combining Eqs. (A1), (A9), and (A11) and expanding in powers of $\delta E$ as in the text gives the same result as Eq. (29) which will not be repeated here. Performing the same integrals then gives

$$|\psi_L'\rangle^{(2)} = (\hat{T}^{-1})^{(2)}|\psi^{(0)}\rangle + (\hat{T}^{-1})^{(1)}|\psi^{(1)}\rangle + (\hat{T}^{-1})^{(0)}|\psi^{(2)}\rangle$$

$$= \frac{d^2 q^2}{2\pi\varepsilon_0 \delta E L^3}\left[1 - \frac{1}{2\pi}\left(\frac{\delta E}{\hbar\omega_L}\right) + \frac{1}{2}\left(\frac{\delta E}{\hbar\omega_A}\right)^2 + ...\right]|0_A\rangle|1_B\rangle|0_F\rangle.$$
$$(A12)$$

Eq. (A12) is the same as the results from the Lorentz gauge in Eq. (30). This shows that the two gauges are indeed related by the transformation $\hat{T}$ and physically equivalent. It is worth noting that this transformation gives the correct results in the Lorentz gauge for each k-vector in the field individually. Thus the equivalence of the two gauges is independent of the results of the integrals, which are primarily useful in showing the dependence on the separation between the two oscillators.

Finally, it has been pointed out that what is traditionally [3-6, 27] referred to as the Lorentz gauge is now sometimes referred to as the Lorenz gauge. The traditional terminology is used here.

**REFERENCES**

1. S. N. Gupta, Proc. Phys. Soc. (London) **A64**, 426 (1950).
2. K. Bleuler, Helv. Phys. Acta. **23**, 567 (1950).
3. S.N. Gupta, *Quantum Electrodynamics* (Gordon and Breach, New York, 1977).
4. S.S. Schweber, *An Introduction to Relativistic Quantum Field Theory* (Harper and Row, New York, 1962).
5. J.M. Jauch and F. Rohrlich, *The Theory of Photons and Electrons* (Springer Verlag, New York, 1976).
6. C. Cohen-Tannoudji, J. Dupont-Roc, and G. Grynberg, *Photons and Atoms: Introduction to Quantum Electrodynamics* (Wiley, New York, 1989).
7. R.M. Gingrich and C. Adami, Phys. Rev. Lett. **89**, 270402 (2002).
8. U. Yurtsever and J.P. Dowling, Phys. Rev. A **65**, 052317 (2002).
9. M. Reisenberger and C. Rovelli, Phys. Rev. D **65**, 125016 (2002).
10. P. Kok, U. Yurtsever, S.L. Braunstein, and J.P. Dowling, arXiv:quant-ph/0206082 (2002).
11. P.M. Alsing and G.J. Milburn, Quant. Inf. Comput. **2**, 487 (2002).
12. A. Peres and D.R. Terno, J. Mod. Opt. **50**, 1165 (2003).
13. A.J. Bergou, R.M. Gingrich, and C. Adami, Phys. Rev. A **68**, 042102 (2003).
14. A. Peres and D.R. Terno, Int. J. Quant. Inf. **1**, 225 (2003).
15. A. Peres and D.R. Terno, Rev. Mod. Phys. **76**, 93 (2004).
16. P. Kok, T.C. Ralph, and G.J. Milburn, Quant. Inf. Comput. **5**, 239 (2005).
17. S.D. Bartlett and D.R. Terno, Phys. Rev. A **71**, 012302 (2005).
18. P. Kok and S.L. Braunstein, Int. J. Quant. Info. **4**, 119 (2006).
19. M. Czachor, Phys. Rev. A **55**, 72 (1997).
20. D.R. Terno, in *Quantum Information Processing: From Theory to Experiment*, D.G. Angelakis et al, ed. (IOP Press, 2006).
21. S.J. Olson and J.P. Dowling, arXiv:0708.3535 (2008).
22. S.J. Olson and T.C. Ralph, Phys. Rev. Lett. **106**, 110404 (2011).

23. T.G. Downes, I. Fuentes, and T.C. Ralph, Phys. Rev. Lett. **106**, 210502 (2011).
24. J.L. Pienaar, C.R. Myers, and T.C. Ralph, Phys. Rev. A **84**, 022315 (2011).
25. S.J. Olson and T.C. Ralph, arXiv:1101.2565 (2011).
26. G. Baym, *Lectures on Quantum Mechanics* (W.A. Benjamin, Reading, 1969).
27. J.D. Jackson, *Classical Electrodynamics* (Wiley, New York, 1967).
28. K. Haller and R.B. Sohn, Phys. Rev. A **20**, 1541 (1979).
29. For a review, see M.B. Plenio and S. Virmani, Quant. Inf. Comp. **7**, **1** (2007).
30. W.K. Wooters, Phys. Rev. Lett. **80**, 2245 (1998).